\documentclass[prb,twocolumn,preprintnumbers,superscriptaddress]{revtex4}

\usepackage{graphicx,epstopdf}
\usepackage{amssymb,amsmath}
\usepackage{dcolumn}
\usepackage{bm}
\usepackage{color}
\usepackage{enumerate}

\begin{document}

\title{Spin currents of exciton polaritons in a microcavity with (110)-oriented quantum well}

\author{V. Shahnazaryan}
\affiliation{Science Institute, University of Iceland, Dunhagi-3, IS-107, Reykjavik, Iceland}
\affiliation{Institute of Mathematics and High Technologies, Russian-Armenian (Slavonic) University, Hovsep Emin 123, 0051, Yerevan, Armenia}

\author{S. Morina}
\affiliation{Science Institute, University of Iceland, Dunhagi-3, IS-107, Reykjavik, Iceland}
\affiliation{Division of Physics and Applied Physics, Nanyang Technological University 637371, Singapore}

\author{S.\,A. Tarasenko}
\affiliation{Ioffe Institute, 194021 St. Petersburg, Russia}
\affiliation{St. Petersburg State Polytechnic University, 195251 St. Petersburg, Russia}

\author{I.\,A. Shelykh}
\affiliation{Science Institute, University of Iceland, Dunhagi-3, IS-107, Reykjavik, Iceland}
\affiliation{Division of Physics and Applied Physics, Nanyang Technological University 637371, Singapore}
\affiliation{ITMO University, St. Petersburg 197101, Russia}

\date{\today}

\begin{abstract}
We study the polarization optical properties of microcavities with embedded (110)-oriented quantum wells. The spin dynamics of exciton polaritons in such structures is governed by the interplay of the spin-orbit splitting of exciton states, which is odd in the in-plane momentum, and the longitudinal-transverse splitting of cavity modes, which is even in the momentum. We demonstrate the generation of polariton spin currents by linearly polarized optical pump and analyze the arising polariton spin textures in the cavity plane.
Tuning the excitation spot size, which controls the polariton distribution in the momentum space, one obtains symmetric
or asymmetric spin textures.
\end{abstract}

\maketitle

\section{Introduction}

The generation and detection of spin currents is among the major research topics of modern spintronics.\cite{Wolf2001} The pure spin current is a combination of flows of particles carrying opposite spin projections in the opposite directions and, ideally, is accompanied by no charge current. In semiconductors with strong spin-orbit (SO) coupling, the pure spin current emerges in the direction perpendicular to a driving charge current. This phenomenon, proposed by D'yakonov and Perel'\cite{Dyakonov1971} and referred to as the spin Hall effect (SHE),\cite{Hirsch1999} is widely studied both theoretically and experimentally.\cite{Kato2004, Wunderlich2005, Sinova2006, Tarasenko2006, Jungwirth2012, Nichele2015} Spin currents can be also optically injected in semiconductor structures lacking the space inversion symmetry at interband,\cite{Bhat2005,Tarasenko2005,Zhao2005,Bieler2006,Duc2010} intersubband,\cite{Tarasenko2005,Sherman2005,Diehl2007,Kyriienko2013} or intrasubband~\cite{Tarasenko2005,Ganichev2006,Olbrich2012} optical transitions. For the direct one-photon absorption in quantum wells (QWs), the effect comes from SOI terms, which are odd in the in-plane wave vector $\bm{k}$, in the electron/hole dispersions and the optical transition rates.\cite{Ivchenko2008}

A difficulty in developing electronic devices operating with spin currents is related to fast relaxation processes in the electron (or hole) system and the rather short corresponding mean free path.\cite{Jungwirth2012} A way to overcome this problem is the development of optical analogues of electronic devices. In spinoptronics,\cite{Shelykh2004} spin currents are carried by exciton-polaritons which are electrically neutral hybrid quasiparticles emerging in semiconductor microcavities with embedded QWs in the regime of strong light-matter coupling.\cite{Microcavities} Polaritons have $\pm1$ spin projections onto the QW growth direction, corresponding to the right- and left-handed circular polarizations, and extremely small effective mass providing mm-scale coherence lengths in high-quality structures.\cite{ShelykhReview} The role of SOI splitting is usually played here by the longitudinal-transverse (LT) splitting of polariton states. The LT splitting originates from the transverse-electric transverse-magnetic splitting of the cavity modes~\cite{Panzarini} and the splitting of exciton states caused by exchange interaction between electrons and heavy holes forming the excitons.\cite{Maialle} LT splitting gives rise to a spin separation of exciton-polaritons injected in a microcavity, known as the optical spin Hall effect (OSHE).\cite{KavokinOSHE,Leyder,Maragkou,Langbein} The spin pattern of OSHE can be modified by an external magnetic field,\cite{Skender} in-plane anisotropy of the microcavity,\cite{Amo,Liew2009,Glazov2010,Dufferwiel} polariton-polariton interactions,\cite{Liew2012,Flayac1}  etc.

The key difference between SO splitting of electron subbands in QWs and LT splitting of polariton branches in microcavities is that
the SO splitting is odd in $\bm k$\cite{Winkler_book} while the LT splitting is even in $\bm k$.\cite{Microcavities} Therefore, in OSHE
polaritons propagating in the opposite in-plane directions carry the same polarization and the resulting spin texture caused by the LT splitting is symmetric.\cite{KavokinOSHE,Leyder,Maragkou,Langbein} The excitation of spin currents of polaritons, where particles with the opposite spin projections counterflow, is a challenging task. Here, we propose a polariton system with both SO and LT interactions. We show that, in a microcavity with (110)-oriented zinc-blende-type QW, the effective Hamiltonian of polaritons contains both
$\bm k$-linear SO and $\bm k$-quadratic LT terms. The $\bm k$-linear SO terms in the polariton Hamiltonian originate from the corresponding $\bm k$-linear Dresselhaus terms in the effective electron and hole Hamiltonians which emerge, in turn, due to low spatial symmetry of (110)-oriented QWs. Note that, in common (001)-oriented structures, $\bm k$-linear terms in the effective electron and hole Hamiltonians are also present but do not lead to $\bm k$-linear terms in the polariton dispersion. Although the unique properties of SO interaction in
(110)-oriented QWs~\cite{Dyakonov,Cartoixa06,Tarasenko2009,Nestoklon2012,Poshakinskiy2013,Wang2014} leading, e.g., to long spin lifetime achievable,~\cite{Ohno1999,Karimov03,Couto2007,Belkov08,Muller08,Griesbeck2012,Voelkl2014} are well known, polariton dynamics in such structures has not been studied so far.

In the present paper, we study the polarization optical properties of the (110)-grown QW embedded into a microcavity. We calculate the dispersion of exciton polaritons taking into account both SO interaction and LT splitting. It is found that, in structures with spin-degenerate polariton states at $\bm k = 0$, the polariton splitting is dominated by $\bm k$-linear SO term at small wave vectors and $\bm k$-quadratic LT term at larger wave vectors. We compute the distribution of polariton spin polarization in real space when polaritons are injected into the cavity by linearly-polarized light at the normal-incidence geometry and show that the resulting spin pattern drastically depends on the lateral size of the excitation spot. For large excitation spots, when polaritons are primarily injected into the states with small $\bm k$, the $\bm k$-linear splitting determines the polariton dynamics and leads to an asymmetric distribution of the polariton spin in real space. For small excitation spots, the polariton dynamics is governed by the $\bm k$-quadratic LT splitting and the symmetric spin pattern of OSHE emerges.


\section{model and theory}

The effective Hamiltonian describing the spin-orbit splitting of the electron and hole subbands in symmetric (110)-oriented QWs to first order in the in-plane wave vector $\bm k$ has the form~\cite{Dyakonov,Winkler_book}
\begin{equation}
H_{so}^{(\nu)} = \gamma_{\nu} \,\tau_z k_x \:,
\end{equation}
where $\nu$ is the subband index, $\tau_z$ is the Pauli matrix in the space of the spin states $\pm 1/2$ for electrons or the space
of the states $\pm 3/2$ for heavy holes, $x \parallel [1\bar{1}0]$ and $y \parallel [00\bar{1}]$ are the in-plane axes, and $z \parallel [110]$ is the growth axis. Accordingly, the dispersion of mechanical excitons with the spin projections $\pm 1$ onto the growth axis formed from electrons and heavy holes is given by
\begin{equation}\label{exc_disp}
E_{X, \pm 1} = E_{X0} + \dfrac{\hbar^2 \bm{k}^2}{2 m_X} \pm \gamma_X k_x \:,
\end{equation}
where $E_{X0}$ is the exciton energy at $\bm k = 0$, $m_X = m_e + m_{hh}$ is the exciton mass, $m_e$ and $m_{hh}$ are the in-plane electron and hole masses,
and
\begin{equation}
\gamma_X = \frac{m_e\gamma_e + m_{hh}\gamma_{hh}}{m_e + m_{hh}} .
\end{equation}
We note that the splitting parameters $\gamma_{\nu}$ drastically depend on the QW width and chemical composition.\cite{Nestoklon2012} Besides the $k_x$-linear term in the exciton dispersion, the spin-orbit coupling may modify the internal structure of excitons leading to the admixture of
excited exciton states to the ground exciton state. The study of the exciton internal structure together with a possible anisotropy of the exciton effective masses in the QW plane, which may arise in (110)-oriented structures, is out of the scope of the present paper.

\begin{figure}[t]
    \includegraphics[width=0.9\linewidth]{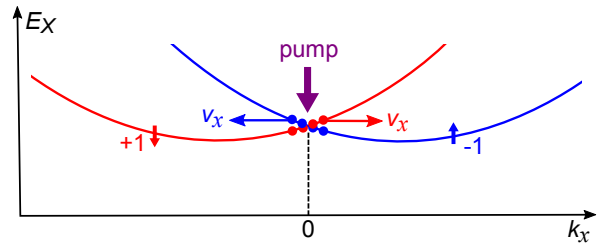}
    \caption{Pure spin current of excitons in (110)-oriented QW. Optical pumping at the normal incidence of light upon the QW generates excitons at $\bm k = 0$. Due to $k_x$-linear terms in the energy spectrum, excitons with the spin projections $\pm 1$ posses non-zero group velocities $v_x$ at $\bm k = 0$ and move in the opposite directions.}
    \label{Fig1}
\end{figure}

The exciton spectrum~\eqref{exc_disp} is shown in Fig.~\ref{Fig1}. The spectrum consists of two spin branches with minima at $k_{x} = \pm \gamma_X m_X /\hbar^2$, respectively. A resonant interband optical excitation of the QW at normal incidence of light generates excitons at $\bm k = 0$, Fig.~\ref{Fig1}. Due to the spin-orbit splitting of the spectrum, excitons with the spin projections $\pm 1$ and the wave vector $\bm k = 0$ have the group velocities $v_x = \pm \gamma_X/\hbar$ and, therefore, move along the $x$ axis in the opposite directions. Circularly polarized light creates excitons with predominant spin projection $+1$ or $-1$ determined by the light helicity. Therefore, the distribution of excitons created by a focused light beam with circular polarization will shift in the QW plane along or
opposite to the $x$ axis depending on the light helicity. Linearly polarized light creates excitons with the spin projection $\pm 1$ at equal rates. In this case, the optical pumping will generate spin currents of excitons where particles with the opposite spin projections counterflow. The resulting spin separation in real space  can be estimated by $\Delta L = 2 (\gamma_X/\hbar) \tau$ which yields $\Delta L \sim 0.3$\,$\mu$m for $\gamma_X = 100$\,meV$\cdot$\AA~\cite{Nestoklon2012} and the relaxation time $\tau = 10$\,ps.
\begin{figure*}[t]
    \includegraphics[width=1.0\linewidth]{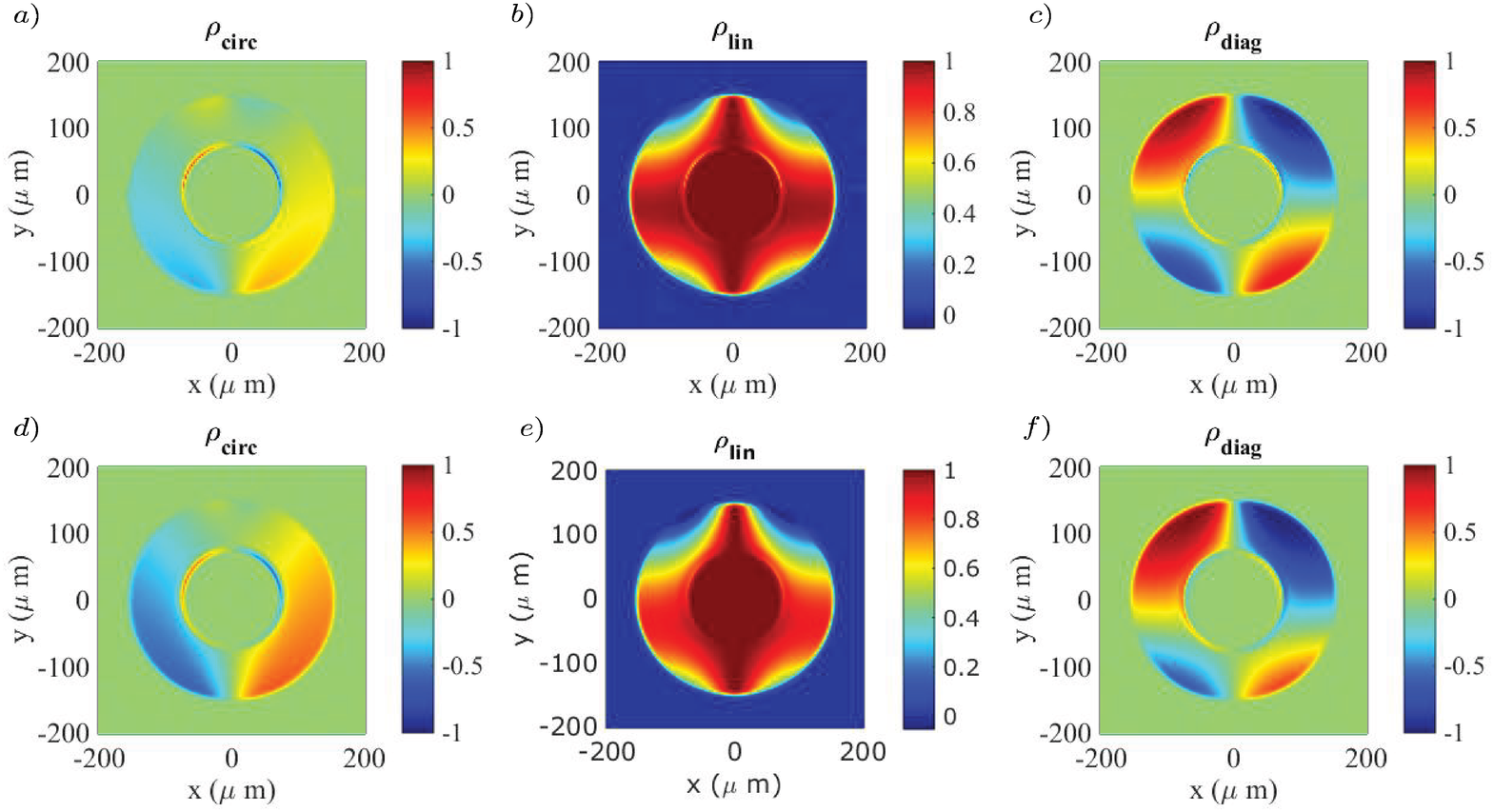}
    \caption{Spatial distribution of the Stokes parameters determining the polarization of polaritons for linearly polarized pump and
				the pump-spot radius $r_0 = 20$\,$\mu$m. Panels a-c and d-f are calculated for the spin-orbit coupling parameters $\gamma_X = 50$\, meV$\cdot$\AA~ and $\gamma_X = 100$\, meV$\cdot$\AA, respectively.}
    \label{Fig2}
\end{figure*}
\begin{figure*}[t]
    \includegraphics[width=1.0\linewidth]{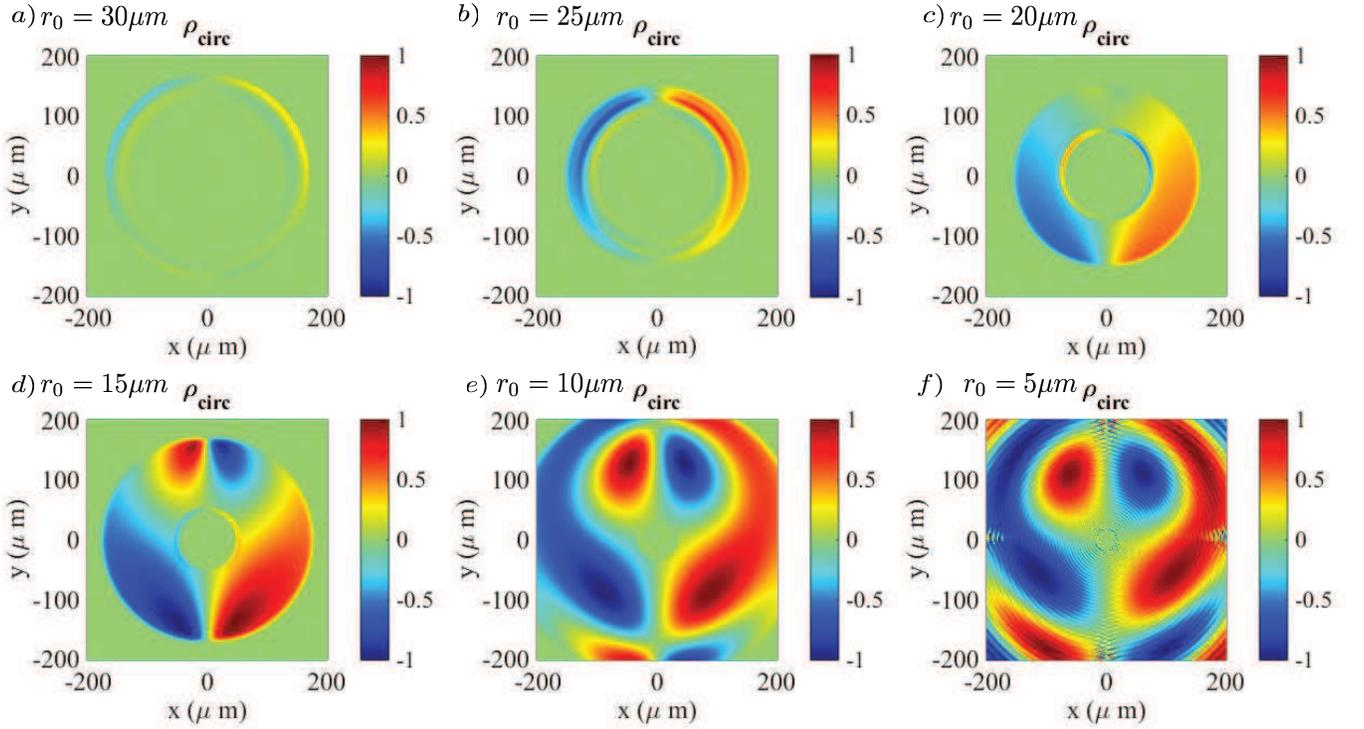}
    \caption{Spatial distribution of the Stoke parameter $\rho_{circ}$ determining the circular polarization of polaritons
		for different pump-spot radii. The pump is linearly polarized along the $x$ axis, $\gamma_X = 100$\, meV$\cdot$\AA.}
    \label{Fig3}
\end{figure*}
\begin{figure*}[t]
    \includegraphics[width=1.0\linewidth]{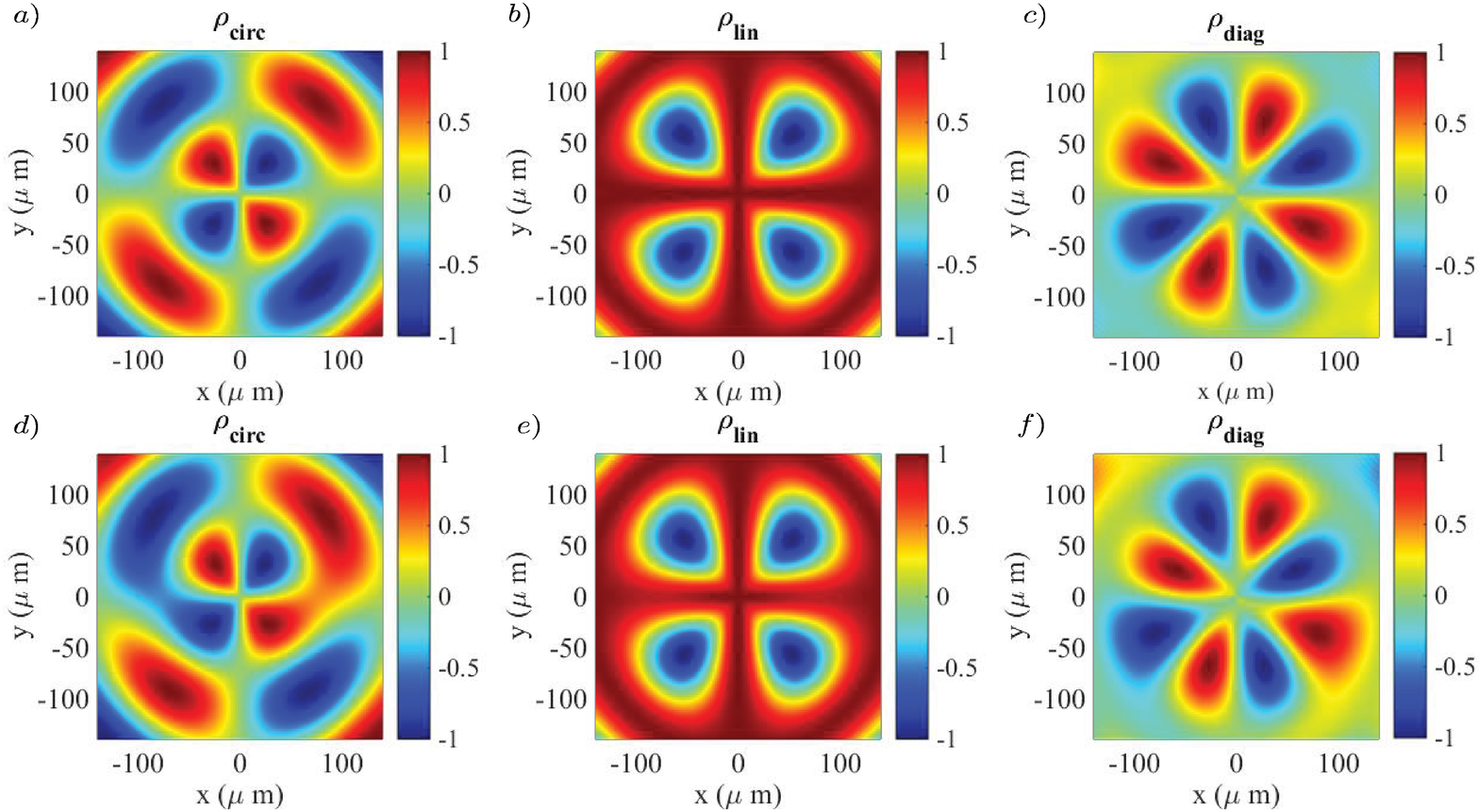}
    \caption{Spatial distribution of the Stokes parameters determining the polarization of polaritons for linearly polarized pump and
				the pump-spot radius  $r_0 = 2$\,$\mu$m. Panels a-c and d-f are calculated for the spin-orbit coupling parameters $\gamma_X = 0$\, meV$\cdot$\AA~ and $\gamma_X = 100$\, meV$\cdot$\AA, respectively.}
    \label{Fig4}
\end{figure*}

To increase the propagation length, exciton may be coupled with an optical mode of the cavity.\cite{Langbein} In such a system, the interaction of cavity photons with QW excitons is drastically enhanced due to multiple reflection of photons from the Bragg mirrors constituting the cavity, which leads to the formation of hybrid exciton-polariton modes. The most general form of the effective Hamiltonian of exciton polaritons can be derived in the framework of the group-representation theory which is based on symmetry arguments. Symmetric QW structures grown from zinc-blende-type compounds along the $[110]$ axis are described by the point group $C_{2v}$ which consists of the two-fold rotation axis $C_2 \parallel [00\bar{1}]$, mirror planes
$M_1 \parallel (1\bar{1}0)$ and $M_2 \parallel (110)$, and the identity element. In such systems, the effective Hamiltonian of exciton polaritons in the lower polariton branch in the basis of circularly polarized components is given by
\begin{equation}\label{H_gen}
H = \left( E_0 + \frac{\hbar^2 k_x^2}{2m_x} + \frac{\hbar^2 k_y^2}{2m_y} \right) I + \gamma k_x \sigma_z
\end{equation}
\[
+ \left( \Delta + \delta_{xx} k_x^2 - \delta_{yy} k_y^2 \right) \sigma_x + 2 \delta_{xy} k_x k_y \sigma_y \:,
\]
where $I$ is the $2 \times 2$ identity matrix, $\sigma_j$ ($j=x,y,z$) are the Pauli matrices, $m_x$ and $m_y$ are the polariton masses, $\gamma$ is the constant of $\bm k$-linear spin-orbit splitting, $\delta_{xx}$, $\delta_{yy}$, and $\delta_{xy}$ are the parameters of longitudinal-transverse (LT) splitting of polaritons, and $\Delta$ describes the splitting of polariton states into two orthogonal linearly polarized components at $\bm k =0$ which can occur in (110)-oriented structures. Hamiltonian~\eqref{H_gen} is valid to second order in the in-plane wave vector. The splitting at $\bm k = 0$ can originate from optical anisotropy of the cavity\cite{Malpuech2006} or
unequal optical transition rates for QW excitons polarized along $x$ and $y$.\cite{Gershoni1991,Nojima1993,Winkler1996,Singh2013} The splitting in (110)-oriented structures can be also tuned by applying strain along the growth axis and made to vanish.

In the present paper, we focus on the spin currents of exciton polaritons caused by SO coupling. Therefore, we consider the structure with $\Delta =0$, isotropic polariton mass, $m_x = m_y$, and isotropic LT splitting, $\gamma_{xx} = \gamma_{yy} = \gamma_{xy}$. In this case, the Hamiltonian can be presented in the form
\begin{equation}\label{H}
H = \left(
\begin{matrix}
E_0 + \dfrac{\hbar^2 k^2}{2m} + \gamma k_x \; & \;  \delta \left(k_x-ik_y\right)^2 \\
\delta\left(k_x+ik_y\right)^2 \; & \; E_0 + \dfrac{\hbar^2 k^2}{2m} - \gamma k_x \\
\end{matrix}
\right).
\end{equation}

The parameters of the effective Hamiltonian~\eqref{H} can be expressed via the corresponding parameters of the QW excitons and cavity photons and the exciton-photon coupling strength. By applying the standard microscopic description of exciton polaritons,\cite{Microcavities}  we obtain
\begin{equation}
E_0 = \frac{1}{2} \left[ E_{C0} + E_{X0} - \sqrt{(E_{C0}-E_{X0})^2 + V^2}  \right] ,
\end{equation}
\begin{equation}
1/m = C/m_C \:, \;\; \gamma = X \gamma_X \:, \;\; \delta = C \delta_C \:,
\end{equation}
where $E_{C0} = E_{X0} - \Delta E$ is the energy of the cavity photon mode at $\bm k =0$, $\Delta E$ is the detuning between the exciton and photon modes, $m_C$ and $\delta_C$ are the cavity photon effective mass and LT splitting, respectively, $V$ is the Rabi splitting energy, and $X$ and $C$ are the excitonic and photonic partitions,
\begin{equation}
    \begin{aligned}
& X = \frac{E_{C0}-E_{X0} + \sqrt{(E_{X0}-E_{C0})^2 + V^2}}{2\sqrt{(E_{X0}-E_{C0})^2 + V^2}} \:, \\
& C = \frac{E_{X0}-E_{C0} + \sqrt{(E_{X0}-E_{C0})^2 + V^2}}{2\sqrt{(E_{X0}-E_{C0})^2 + V^2}} \:. \\
    \end{aligned}
\end{equation}
The LT splitting of QW excitons is neglected since it is typically three orders of magnitude weaker than that of the cavity photons.\cite{Flayac3}

To describe the dynamics of exciton polaritons we solve the Schr\"{o}dinger-type equation accounting for the external pump and decay
\begin{equation}\label{Schroedinger}
i\hbar\frac{\partial\Psi}{\partial t} = H \Psi - \frac{i\hbar}{2\tau}\Psi + P(\bm r,t)
\end{equation}
for the two-component wave function
\begin{equation}\label{Psi}
\Psi=
\left(
\begin{matrix}
\psi_+(\bm r,t) \\
\psi_-(\bm r,t) \\
\end{matrix}
\right) ,
\end{equation}
where $\psi_+$ and  $\psi_-$ stand for the right-handed and left-handed circularly polarized components, respectively. The second term on the right-hand side of Eq.~\eqref{Schroedinger} describes the decay of polaritons with the lifetime $\tau$ while the last term stands for the external pump. We assume that the pump beam is normal to the QW plane, linearly polarized and we take the profile $P(\bm{r},t)$ in the form of a Gaussian spot, namely
\begin{equation}
P(\bm r,t)=P_0 \exp{\left( - i \frac{E_0 + E_P}{\hbar} t\right)} \exp{\left( - \frac{\bm r^2}{r_0^2} \right)} \left(\begin{matrix} 1\\1\\\end{matrix}\right),
\end{equation}
where $E_P$ is the injected polariton energy measured from the minimum of the lower polariton branch and $r_0$ is the pump spot radius.

The wave function~\eqref{Psi} straightforwardly determines the Stokes parameters of the emitted light
\begin{equation}
    \begin{aligned}
\rho_{circ} &=\frac{|\psi_+|^2-|\psi_-|^2}{|\psi_+|^2+|\psi_-|^2} \:,\\
\rho_{lin} & =\frac{\psi_+^*\psi_-+\psi_-^*\psi_+}{|\psi_+|^2+|\psi_-|^2} \:,\\
\rho_{diag} & =i\frac{\psi_+^*\psi_--\psi_-^*\psi_+}{|\psi_+|^2+|\psi_-|^2} \:,\\
    \end{aligned}
\end{equation}
which correspond to the circular, linear, and diagonal polarization degrees, respectively.

\section{Results and Discussion}

We simulate the spin dynamics of exciton polaritons by solving Eq.~\eqref{Schroedinger} numerically for the parameters corresponding to
GaAs/AlGaAs heterostructures. We take the polariton lifetime to be $\tau=10$\,ps, the strength of the Rabi splitting $V=5$\,mev, the detuning between the cavity photon mode and the exciton mode $\Delta E=-0.5$\,mev, and the in-plane mass of the cavity photons $m_C=5*10^{-5}m_0$. The constant of the photon LT splitting is given by $\delta_C = \hbar^2/4 (1/m_C^{TM} - 1/m_C^{TE})$, where $m_C^{TE}=m_C$ and $m_C^{TM}=0.975m_C$ are the in-plane masses of the cavity photon TE and TM modes.\cite{Flayac3} The access energy of excited polaritons is chosen to be $E_P=1$\,meV.
The pump is linearly polarized along the $x$ axis. To calculate the steady-state distribution of polaritons in real space, we turn on the pump at $t=0$ and wait approximately $20 \tau$ until the steady state is achieved.

Figure~\ref{Fig2} shows the spatial distribution of polariton polarization given by the Stokes parameters $\rho_{circ}$, $\rho_{lin}$, and $\rho_{diag}$ obtained for a large pump-spot radius $r_0 = 20$\,$\mu$m. The upper and lower panels contain plots for two different values of the spin-orbit coupling parameter $\gamma_X$ which can be tuned in a wide range by changing the QW width. One can see that the pumping with linearly polarized light results in the emergence of a partial circular polarization of polaritons at the right and left sides of the pump spot, Fig.~\ref{Fig2}a and \ref{Fig2}d. The circular polarization $\rho_{circ}$ is an odd function of the $x$ coordinate and originates from the polariton spin current where the ``spin-up'' and ``spin-down'' particles move in the opposite directions. The spin current and the subsequent spin separation in real space is more pronounced for strong spin-orbit coupling, cf. Fig.~\ref{Fig2}a and \ref{Fig2}d.

The patterns of the linear and diagonal polarization components are shown in~Fig.~\ref{Fig2}b,e and~Fig.~\ref{Fig2}c,f, respectively.
The linear polarization of cavity polaritons stems from the linear polarization of the pump light. The diagonal component results from
quantum beats which emerge because the polaritons are excited in an superposition of eigenstates with different energies. In our particular case, the beats emerge because the pump with a finite spot radius $r_0$ excites polaritons with the wave vectors $k \sim 1/r_0$ and the polarization of polariton eigenstates at $k \neq 0$ does not coincide generally with the linear polarization of the pump. The quantum beats can be instructively described in the framework of the pseudospin formalism as a precession of the polariton pseudospin in an effective magnetic field.\cite{KKavokin} The Larmor frequency corresponding to the effective field is given by the expansion
of the Hamiltonian~\eqref{H} in the Pauli matrices and has the form
\begin{equation}
\Omega_x = \frac{2 \delta (k_x^2-k_y^2)}{\hbar} \,, \;\;
\Omega_y = \frac{4 \delta  k_x k_y}{\hbar} \,, \;\; \Omega_z = \frac{2 \gamma k_x}{\hbar} \,.
\end{equation}
The effective field is determined by both LT splitting and SO splitting and, therefore, contains both even-in-$\bm k$ and odd-in-$\bm k$ contributions. The polariton pseudospins initially pointed along the $x$ axis precess in the effective field which leads to the
emergence of the diagonal polarization component. At large spot sizes (when only states with small wave vectors are populated) and strong SO coupling, the $k_x$-linear contribution to the effective field plays a major role and the pattern of $\rho_{diag}$ is asymmetric, see Fig.~\ref{Fig2}f.

A decrease in the pump-spot size leads to the population of polariton states with higher wave vectors. As a consequence, the $\bm k$-quadratic LT splitting takes over the $\bm k$-linear SO splitting and the spatial distribution of polarization changes. Such a transformationof the pattern of the polariton circular polarization is shown in Figure~\ref{Fig3}. At large radii of the pump spot,
the spatial distribution of $\rho_{circ}$ is asymmetric and originates from the spin current of polaritons. At smaller radii,
the asymmetric distribution is superimposed with the symmetric cross-shape pattern of the optical spin Hall effect stemming
from the LT splitting.\cite{KavokinOSHE}

Finally, we discuss the spatial distribution of the polariton polarization in the case of a small pump-spot radius, Fig.~\ref{Fig4}.
In this geometry, polaritons are excited into the states with relatively high wave vectors and their spin dynamics is mostly governed by the LT splitting leading to the symmetric patterns of the OSHE. Still, a considerable distortion of the spatially symmetric distribution of $\rho_{circ}$ and the emergence of an asymmetric component of the distribution in the presence of SO splitting are clearly seen, fc. Fig.~\ref{Fig4}a and Fig.~\ref{Fig4}d. In the pictures of linear and diagonal polarizations, the influence of the SOI field is not so pronounced, manifesting itself in the slight squeezing of linear polarization pattern and appearance of minor asymmetry in the diagonal polarization pattern (see fc. Fig.~\ref{Fig4}b and Fig.~\ref{Fig4}e, and fc. Fig.~\ref{Fig4}c and Fig.~\ref{Fig4}f respectively).

\section{Summary}

We have theoretically studied the generation of spin currents of exciton polaritons in a microcavity with (110)-oriented quantum well. It has been shown that the pattern of spin currents in such structures is caused by the interplay of the spin-orbit splitting of the polariton branch, which is linear in the in-plane momentum, and the longitudinal-transverse splitting, which is quadratic in the momentum.
Taking GaAs/AlGaAs structure as an example, we have analyzed the arising polariton spin textures in the cavity plane, described by the spatially-resolved Stokes parameters, when polaritons are excited by focused linearly polarized light. If the light beam is wide, polaritons are primarily injected into the states with small in-plane momenta where the spin-orbit splitting dominates. The spin-orbit splitting leads to the generation of spin currents, where polaritons with opposite spin projections flow in the opposite directions, resulting in an asymmetric spin separation in real space. For narrow light beams, when polaritons populate states with large wave vectors, the longitudinal-transverse splitting determines the polariton spin dynamics which results in the formation of the OSHE symmetric spin texture.

\section*{Acknowledgements}

We thank Dr. O. Kyriienko for useful discussions. The work was partly supported by FP7 ITN NOTEDEV, Rannis project BOFEHYSS, Russian
Foundation for Basic Research, and the RF President Grant MD-3098.2014.2.



\begin{thebibliography}{99}

\bibitem{Wolf2001} S. A. Wolf, D. D. Awschalom, R. A. Buhrman, J. M. Daughton, S. von Molna´r, M. L. Roukes, A. Y. Chtchelkanova, and D. M. Treger, Science \textbf{294}, 1488 (2001).


\bibitem{Dyakonov1971} M. I. D'yakonov and V. I. Perel', Pis'ma Zh. Eksp. Teor. Fiz. {\bf 13}, 657 (1971)
[JETP Lett. {\bf 13}, 467 (1971)].

\bibitem{Hirsch1999} J.E. Hirsch, Phys. Rev. Lett. \textbf{83}, 1834 (1999).

\bibitem{Kato2004} Y. K. Kato, R. C. Myers, A. C. Gossard, and D. D. Awschalom,
Science {\bf 306}, 1910 (2004).

\bibitem{Wunderlich2005} J. Wunderlich, B. Kaestner, J. Sinova, and T. Jungwirth,
Phys. Rev. Lett. {\bf 94}, 047204 (2005).

\bibitem{Sinova2006} J. Sinova, S. Murakami, S.-Q. Shen, and M.-S. Choi,
Solid State Commun. {\bf 138}, 214 (2006).

\bibitem{Tarasenko2006} S. A. Tarasenko,
Pis'ma Zh. Eksp. Teor. Fiz. {\bf 84}, 233 (2006) [JETP Lett. {\bf 84}, 199 (2006)].

\bibitem{Jungwirth2012} T. Jungwirth, J. Wunderlich, and K. Olejnik,
Nature Mat. {\bf 11}, 382 (2012).

%
%


\bibitem{Nichele2015} F. Nichele, S. Hennel, P. Pietsch, W. Wegscheider, P. Stano, P. Jacquod, T. Ihn, and K. Ensslin, Phys. Rev. Lett. \textbf{114}, 206601 (2015).


\bibitem{Bhat2005} R. D. R Bhat, F. Nastos, Ali Najmaie, and J. E. Sipe,
Phys. Rev. Lett. \textbf{94}, 096603 (2005).

\bibitem{Tarasenko2005} S. A. Tarasenko and E. L. Ivchenko,
Pis'ma Zh. Eksp. Teor. Fiz. {\bf 81}, 292 (2005)
[JETP Lett. \textbf{81}, 231 (2005)].

\bibitem{Zhao2005} H. Zhao, X. Pan, A. L. Smirl, R. D. R. Bhat, A. Najmaie, J. E. Sipe, and H. M. van Driel,
Phys. Rev. B \textbf{72}, 201302 (2005).


\bibitem{Bieler2006} M. Bieler, K. Pierz, U. Siegner, and P. Dawson,
Phys. Rev. B {\bf 73}, 241312(R) (2006).

\bibitem{Duc2010} H. T. Duc, J. F\"{o}rstner, and T. Meier,
Phys. Rev. B {\bf 82}, 115316 (2010).

\bibitem{Sherman2005} E. Ya. Sherman, A. Najmaie, and J. E. Sipe,
Appl. Phys. Lett. {\bf 86}, 122103 (2005).

\bibitem{Diehl2007} H. Diehl, V. A. Shalygin, V. V. Bel'kov, Ch. Hoffmann, S. N. Danilov, T. Herrle, S. A. Tarasenko, D. Schuh,
Ch. Gerl, W. Wegscheider, W. Prettl, and S. D. Ganichev,
New J. Phys. {\bf 9}, 349 (2007).

\bibitem{Kyriienko2013} O. Kyriienko and I. A. Shelykh, Phys. Rev. B \textbf{87}, 075446 (2013).

\bibitem{Ganichev2006} S. D. Ganichev, V. V. Bel'kov, S. A. Tarasenko, S. N. Danilov, S. Giglberger, Ch. Hoffmann, E. L. Ivchenko, D. Weiss, W. Wegscheider, C. Gerl, D. Schuh, J. Stahl, J. De Boeck, G. Borghs, and W. Prettl,
Nature Phys. {\bf 2}, 609 (2006).

\bibitem{Olbrich2012} P. Olbrich, C. Zoth, P. Lutz, C. Drexler, V.V. Bel'kov, Ya.V. Terent'ev, S.A. Tarasenko, A.N. Semenov, S.V. Ivanov, D.R. Yakovlev, T. Wojtowicz, U. Wurstbauer, D. Schuh, and S.D. Ganichev,
Phys. Rev. B {\bf 86}, 085310 (2012).

\bibitem{Ivchenko2008} E. L. Ivchenko and S. A. Tarasenko,
Semicond. Sci. Technol. {\bf 23}, 114007 (2008).


\bibitem{Shelykh2004} I.A. Shelykh, K.V. Kavokin, A.V. Kavokin, G. Malpuech, P. Bigenwald, H. Deng, G. Weihs, Y. Yamamoto, Phys. Rev. B \textbf{70}, 035320 (2004).

\bibitem{Microcavities} A. Kavokin, J. J. Baumberg, G. Malpuech, and F. P. Laussy, \textit{Microcavities} (Oxford University Press, New York, 2007).

\bibitem{ShelykhReview} I.A. Shelykh, A. V. Kavokin, Y. G. Rubo, T. C. H. Liew and G. Malpuech, Semicond. Sci. Technol. \textbf{25}, 013001 (2010).

\bibitem{Panzarini} G. Panzarini, L.C. Andreani, A. Armitage, D. Baxter, M.S. Skolnick, V.N. Astratov, J.S. Roberts, A.V. Kavokin, M.R. Vladimirova and M.A. Kaliteevski, Phys Rev B \textbf{59}, 5082 (1999).

\bibitem{Maialle} M. Z. Maialle, E. A. de Andrada e Silva and L. J. Sham, Phys. Rev. B \textbf{47}, 15776 (1993).

\bibitem{KavokinOSHE} A. Kavokin, G. Malpuech, M. Glazov, Phys. Rev. Lett. \textbf{95}, 136601 (2005).

\bibitem{Leyder} C. Leyder, M. Romanelli, J. Ph. Karr, E. Giacobino, T. C. H. Liew, M. M. Glazov, A. V. Kavokin, G. Malpuech, and A. Bramati, Nature Phys. \textbf{3}, 628 (2007).

\bibitem{Maragkou} M. Maragkou, C.E. Richards, T. Ostatnický, A.J.D. Grundy, J. Zajac, M. Hugues, W. Langbein, and P.G. Lagoudakis, Optics Letters \textbf{36}, 1095-1097 (2011).

\bibitem{Langbein} W. Langbein, I.A. Shelykh, D. Solnyshkov, G. Malpuech, Yu. Rubo and A. Kavokin, Phys. Rev. B \textbf{75}, 075323 (2007).

\bibitem{Skender} S. Morina, T. C. H. Liew, and I. A. Shelykh, Phys. Rev. B \textbf{88}, 035311 (2013).

\bibitem{Amo} A. Amo, T.C.H. Liew, C. Adrados, E. Giacobino, A. V. Kavokin, and A. Bramati, Phys. Rev. B \textbf{80}, 165325 (2009).

\bibitem{Liew2009} T. C. H. Liew, C. Leyder, A. V. Kavokin, A. Amo, J. Lefrere, E. Giacobino, and A. Bramati Phys. Rev. B \textbf{79}, 125314 (2009).

\bibitem{Glazov2010} M. M. Glazov and L. E. Golub,
Phys. Rev. B {\bf 82}, 085315 (2010).

\bibitem{Dufferwiel} S. Dufferwiel, Feng Li, E. Cancellieri, L. Giriunas, A. A. P. Trichet, D. M. Whittaker, P. M. Walker, F. Fras, E. Clarke, J. M. Smith, M. S. Skolnick, and D. N. Krizhanovskii1, arXiv:1504.02341v1 (2015).
		
\bibitem{Liew2012} E. Kammann, T.C.H. Liew, H. Ohadi, P. Cilibrizzi, P. Tsotsis, Z. Hatzopoulos, P.G. Savvidis, A.V. Kavokin, and P.G. Lagoudakis, Phys. Rev. Lett. \textbf{109}, 036404 (2012).

\bibitem{Flayac1} H. Flayac, D.D. Solnyshkov, I.A. Shelykh, and G. Malpuech, Phys. Rev. Lett. \textbf{110}, 016404 (2013).


\bibitem{Winkler_book} R. Winkler, {\it Spin-Orbit Coupling Effects in Two-Dimensional Electron and Hole Systems}
(Springer, Berlin, 2003).


\bibitem{Dyakonov} M.I. D'yakonov and V.Y. Kachorovskii, Fiz. Tekh. Poluprovodn. \textbf{20}, 178 (1986)
[Sov. Phys. Semicond. \textbf{20}, 110 (1986)].

\bibitem{Cartoixa06} X.~Cartoix\`{a}, L.-W.~Wang, D.~Z.-Y.~Ting, and Y.-C.~Chang,
Phys. Rev. B {\bf 73}, 205341 (2006).

\bibitem{Tarasenko2009} S.A. Tarasenko,
Phys. Rev. B {\bf 80}, 165317 (2009).

\bibitem{Nestoklon2012} M.O. Nestoklon, S.A. Tarasenko, J.-M. Jancu, and P. Voisin,
Phys. Rev. B {\bf 85}, 205307 (2012).

\bibitem{Poshakinskiy2013}	A.V. Poshakinskiy and S.A. Tarasenko,
Phys. Rev. B {\bf 87}, 235301 (2013).

\bibitem{Wang2014}	G. Wang, A. Balocchi, A.V. Poshakinskiy, C.R. Zhu, S.A. Tarasenko, T. Amand, B.L. Liu, X. Marie,
New J. Phys. {\bf 16}, 045008  (2014).

\bibitem{Ohno1999} Y. Ohno, R. Terauchi, T. Adachi, F. Matsukura, and H. Ohno,
Phys. Rev. Lett. {\bf 83}, 4196 (1999).

\bibitem{Karimov03} O.~Z.~Karimov, G.~H.~John, and R.~T.~Harley, W.~H.~Lau, M.~E.~Flatt\'{e}, M.~Henini, and R.~Airey,
Phys. Rev. Lett. {\bf 91}, 246601 (2003).

\bibitem{Couto2007} O. D. D. Couto, F. Iikawa, J. Rudolph, R. Hey, and P. V. Santos,
Phys. Rev. Lett. {\bf 98}, 036603 (2007).

\bibitem{Belkov08}  V.~V.~Bel'kov, P.~Olbrich, S.~A.~Tarasenko, D.~Schuh, W.~Wegscheider, T.~Korn, Ch.~Schuller, D.~Weiss, W.~Prettl, and S.~D.~Ganichev,
Phys. Rev. Lett. {\bf 100}, 176806 (2008).

\bibitem{Muller08} G.~M.~M\"{u}ller, M.~R\"{o}mer, D.~Schuh, W.~Wegscheider, J.~H\"{u}bner, and M.~Oestreich,
Phys. Rev. Lett. {\bf 101}, 206601 (2008).

\bibitem{Griesbeck2012} M. Griesbeck, M. M. Glazov, E. Y. Sherman, D. Schuh, W.Wegscheider, C. Sch\"{u}ller, and T. Korn,
Phys. Rev. B {\bf 85}, 085313 (2012).

\bibitem{Voelkl2014} R. V\"{o}lkl, M. Schwemmer, M. Griesbeck, S. A. Tarasenko, D. Schuh, W. Wegscheider, C. Sch\"{u}ller, T. Korn,
Phys. Rev. B 89, 075424 (2014).


\bibitem{Malpuech2006} G. Malpuech, M.M. Glazov, I.A. Shelykh, K.V. Kavokin and P. Bigenwald,
Appl. Phys. Lett. \textbf{88}, 111118 (2006)

\bibitem{Gershoni1991} D. Gershoni, I. Brener, G. A. Baraff, S. N. G. Chu, L. N. Pfeiffer, and K. West,
Phys. Rev. B {\bf 44}, 1930(R) (1991).

\bibitem{Nojima1993} S. Nojima,
Phys. Rev. B {\bf 47}, 13535 (1993).

\bibitem{Winkler1996} R. Winkler and A. I. Nesvizhskii,
Phys. Rev. B {\bf 53}, 9984 (1996).

\bibitem{Singh2013} R. Singh, T. M. Autry, G. Nardin, G. Moody, H. Li, K. Pierz, M. Bieler, and S. T. Cundiff,
Phys. Rev. B {\bf 88}, 045304 (2013).


\bibitem{Flayac3} H. Flayac, D.D. Solnyshkov, G. Malpuech, and I.A. Shelykh,
Phys. Rev. B {\bf 87}, 075316 (2013).

\bibitem{KKavokin} K.V. Kavokin, I.A. Shelykh, A.V. Kavokin, G. Malpuech, and P. Bigenwald,
Phys. Rev. Lett. \textbf{92}, 017401 (2004).

\end{thebibliography}
\end{document}